\documentclass[]{aastex631}

\graphicspath{{.}}
\usepackage{graphicx}
\usepackage{multirow}

\begin{document}

\title{The Dependence of Joy's Law and Mean Tilt as a Function of Flux Emergence Phase}

\author{Lucy W. Will}
\affiliation{HEPL Solar Physics, Stanford University, Stanford, CA 94305-4085}

\author[0000-0003-2622-7310]{Aimee A. Norton}
\affiliation{HEPL Solar Physics, Stanford University, Stanford, CA 94305-4085}

\author[0000-0001-9130-7312]{Jon Todd Hoeksema}
\affiliation{HEPL Solar Physics, Stanford University, Stanford, CA 94305-4085}

\begin{abstract}

Data from the Michelson Doppler Imager (MDI) and Helioseismic and Magnetic Imager (HMI) are analyzed from 1996 to 2023 to investigate tilt angles ($\gamma$) of bipolar magnetic regions and Joy's Law for Cycles 23, 24, and a portion of 25. The HMI radial magnetic field ($B_{r}$) and MDI magnetogram ($B_{los}$) data are used to calculate ($\gamma$) using the flux-weighted centroids of the positive and negative polarities. Each active region (AR) is only sampled once. The analysis includes only Beta ($\beta$)-class active regions since computing $\gamma$ of complex active regions is less meaningful.  
During the emergence of the ARs, we find that the average tilt angle ($\bar{\gamma}$) increases from 3.30$^{\circ}\pm$0.75 when 20\% of the flux has emerged to 6.79$^{\circ}\pm$0.66 when the ARs are at their maximum flux. Cycle 24 had a larger average tilt $\bar{\gamma}_{24}$=6.67$\pm$0.66 than Cycle 23, $\bar{\gamma}_{23}$=5.11$\pm$0.61. No significant difference is found in the slope of Joy's law or $\bar{\gamma}$ when sampling the ARs at the time of maximum flux or central meridian crossing. There are persistent differences in $\bar{\gamma}$ in the hemispheres, with the southern hemisphere having higher ${\bar{\gamma}}$ in Cycles 23 and 24, but the uncertainties are such that these differences are not statistically significant. 
\end{abstract}


\section{Introduction}

\citet{hale:1908} showed that  sunspots are formed by strong magnetic fields. Hale's polarity law states that sunspots appear in pairs, with a positive and negative magnetic polarity, and that the magnetic polarity of the sunspot leading with respect to rotation is opposite across the equator. The leading sunspot polarity in each hemisphere switches from one cycle to the next. The implications are that the large-scale organization of the magnetic field in the interior is mostly East-West in orientation and the fields are oppositely directed on either side of the equator. 

On average, bipolar sunspot pairs are oriented such that the leading sunspot (with respect to rotation) in each hemisphere is closer to the equator than the following sunspot \citep{hale:1919}. This orientation is referred to as the tilt angle and is a measure of the orientation of the bipolar magnetic region's axis with respect to a line of constant latitude. The tilt angles increase (becoming more North-South oriented and less East-West oriented) with latitude, and this trend was named ``Joy's Law" by \citet{zirin:1988}.  Different definitions for the zero point and allowed ranges in tilt values exist; a topic discussed at the end of this section. 

Tilt angles are an important aspect of flux-transport dynamo models because the tilt plays a role in the formation and evolution of polar fields [see, e.g., \citet{wang:1991, dikpati:1999}]. Tilt angle scatter has been shown to add variability to solar cycle amplitudes \citep{karak:2017, karak:2023}.  The conversion of toroidal magnetic field into poloidal, i.e., the $\alpha$-effect, and subsequently the reversal of the axial dipole between cycles, can be observed directly via the flux transport of tilted bipolar active regions on the solar surface \citep{cameron:2018}. 

There are several proposed mechanisms for the origin of Joy's law.  \citet{babcock:1961} proposed that the tilt angle reflects the global magnetic field components at depth and is a consequence of the ``winding up" of the poloidal field in the solar interior. \citet{wang:1991} proposed that the Coriolis effect acting on flows within the flux tube as it rises through the convection zone is the cause of Joy's law. However, the findings reported by \citet{schunker:2020} show that the motions of the bipolar magnetic regions were an inherent north-south separation speed of the polarities, dependent on latitude but independent of flux. Their results
indicated that the flows in the flux tube need to be directed away from the loop apex if the Coriolis effect were the cause of Joy's law.

Joy's law is only obvious after averaging and, as such, is a statistical law.  \citet{wang:1989} conducted a study with over 2500 bipolar magnetic regions and reported that 16.6\% had no measurable tilts, 19\,\% were anti-Joy (i.e., the following spot was closer to the equator than the leading spot), and 4.4\,\% were anti-Hale (i.e., the polarity of the leading spot was opposite to the majority of leading spots in that hemisphere for that cycle). That is 39.9\% of regions did not obey Joy's law. In another study by \citet{mcclintock:2013}, the noise inherent in the data is so high that Joy's law cannot be recovered for Cycle 17 in the northern hemisphere and Cycle 19 in the southern hemisphere. The scatter in the tilt angles is thought to have a physical origin -- the buffeting of flux tubes by convective motions \citep{fisher:1995, weber:2011}.  Weaker ARs have higher scatter than stronger ARs \citep{wang:1989}. 

The definition of $\gamma$, the data product used for measurement, and the allowed range of $\gamma$ affect the resultant distribution of measured tilts and any fits to the data. For example, some studies (see \citealt{hale:1919, howard:1991, fisher:1995, dasi-espuig:2010, mcclintock:2013}) use white-light data without polarity information and a limited range of $\gamma\leq\pm90^\circ$. Under these circumstances, the magnetic polarity is unknown, no anti-Hale angles are possible, and therefore many of the recorded $\gamma$ values are incorrect. While data analysis using recorded magnetic field polarities are preferred (see data analysis and results from \citealt{wang:1989, howard:1991b, norton:2005, li:2012, mcclintock:2014, Li:2018, munoz-jaramillo:2021}), white-light data catalogs form the longest, most continuous records and allow research on $\gamma$ for many solar cycles.   

The frequency of sampling and practice of binning data into latitude bins also affects the results. If all ARs on the disk are sampled multiple times a day or daily, the results are biased to be representative of longer-lived ARs. Sampling the same ARs multiple times (as done by \citealt{howard:1991, howard:1991b, fisher:1995, stenflo:2012, dasi-espuig:2010, mcclintock:2013}) increases the sampling size, thus reducing the standard error of the sample, but this is misleading; if the same AR tilt angle is being recorded multiple times, then it is not an independent data point.  

Inconsistency in Joy's law studies are also due to different practices in fitting.  One issue is whether or not the fits are allowed to have a y-intercept or are forced through the origin. The average tilt values near the equator are not zero, suggesting that a fit through the origin may not be warranted. \citet{wang:1989, norton:2005, mcclintock:2013}, and \citet{Li:2018} do not force the fit through 0 and \citet{tlatova:2018} states ``The presence of an offset in the non-zero tilt at solar equator is a clear indication that the Coriolis force alone cannot explain the active region tilt." 

Another issue is the practice of fitting the averages of data binned in latitude as done by \citet{dasi-espuig:2010, stenflo:2012, mcclintock:2013} and \citet{sreedevi:2024}, as opposed to fitting all the data points at once as done by \citet{li:2012, Li:2018}, and this paper. Fitting the binned averages reduces the uncertainty on the returned slope of Joy's law but may misrepresent the data, as the fit gives equal weight to bins containing different numbers of data points. 

Analyzing the tilt angles independently by hemisphere is motivated by observations that the northern and southern hemispheres appear to be only moderately to strongly coupled, producing different sunspot numbers and sunspot areas \citep{temmer:2006} in each cycle and having temporal phase shifts for the peak time of the sunspot production and polar field reversal (see \citealt{norton:2014} and references therein). \citet{li:2012, mcclintock:2013}, and \citet{Li:2018} report differences in $\bar{\gamma}$ in the northern and southern hemispheres that are statistically significant. 


The remainder of the report is structured as follows. Section~\ref{sec:methods} describes the data and methods. Results are presented in Section~\ref{sec:results}, beginning with the average tilt angles and Joy's law characteristics determined for Cycle 24 combining data from both hemispheres. 
Subsections provide additional details: Section~\ref{sec:hem} shows the northern and southern hemispheric data separately. Section~\ref{sec:phase} compares tilts in Cycle 24 when the ARs are 20\% and 100\% emerged to investigate dependence on the phase of emergence. Section~\ref{sec:size} looks at tilt as a function of AR size at the time when the ARs are 100\% emerged. Section~\ref{sec:cycles} explores the differences and similarities between Cycles 23, 24 and 25. Finally, Section~\ref{sec:cm} examines the differences in tilts and Joy's law when ARs are sampled at their central meridian crossing time versus the time when they are 100\% emerged. 
Section~\ref{sec:discussion} discusses the relevance of our findings.  

\begin{figure}
    \centering
    \includegraphics[width=0.8\textwidth]{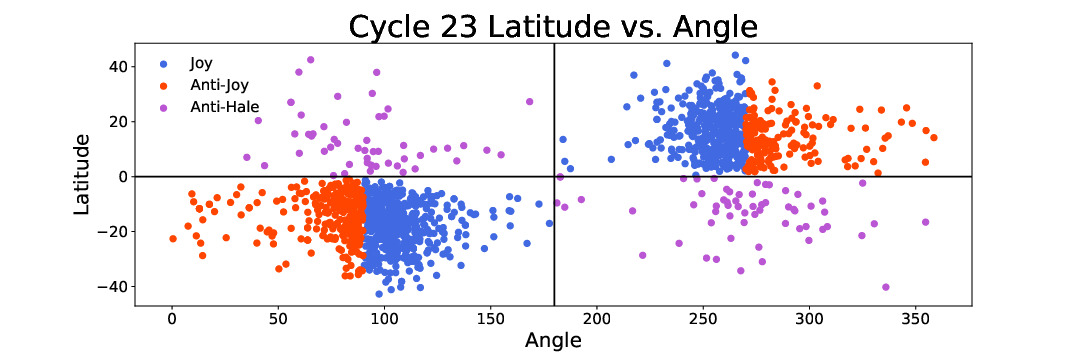}
    \includegraphics[width=.8\textwidth]{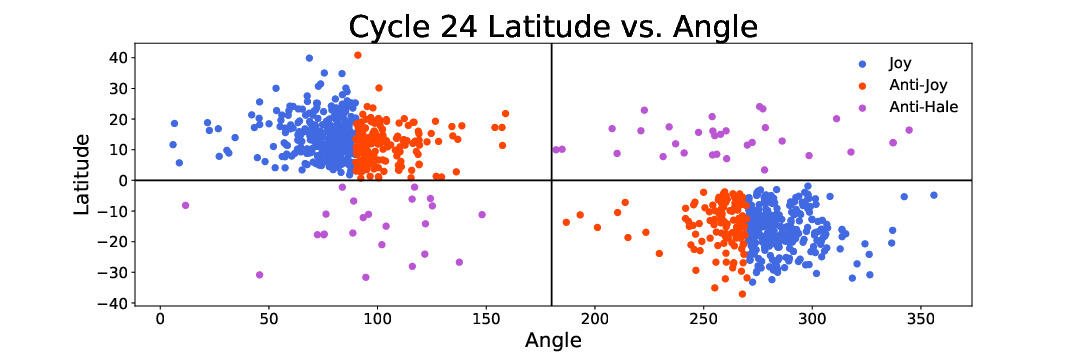}
    \includegraphics[width=0.8\textwidth]{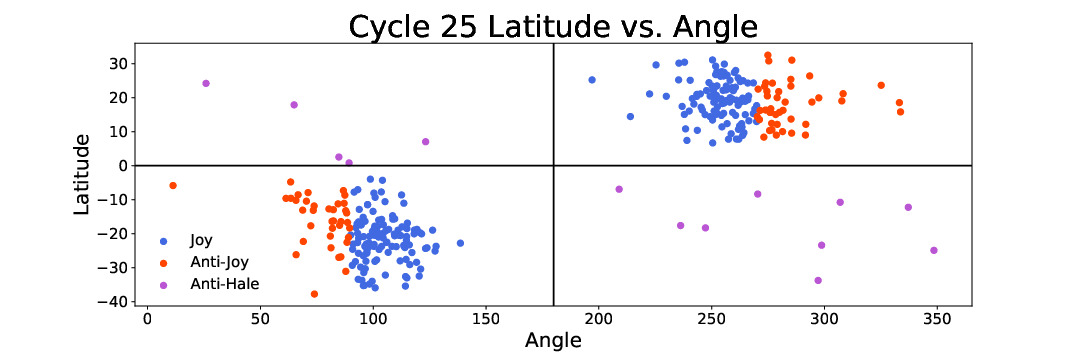}
    \caption{Tilt angle versus latitude for all $\beta$-regions from Cycles 23, 24 and part of 25 with blue/\-red/\-lavender color coding showing Joy/\-anti-Joy/\-anti-Hale regions. $\gamma$ is calculated for a $0-360^\circ$ range defined with the negative polarity always at the origin and a 0$^\circ$  tilt indicating a completely vertical active region with the negative polarity to the north.  Angles increase counter-clockwise. A 90$^\circ$ tilt indicates a completely horizontal AR with negative leading polarity.We exclude from the rest of our analysis all anti-Hale regions, e.g. those data points shown in lavender.}
    \label{fig:tilt-vs-lat}
\end{figure}

\section{Data and Methods}
\label{sec:methods}
We use magnetic data from MDI \citep{scherrer:1995} onboard the Solar and Heliospheric Observatory (SOHO) and HMI \citep{scherrer:2012,Schou:2012} onboard the Solar Dynamics Observatory (SDO) to analyze all beta-type regions during Cycle 23, Cycle 24, and the early stages of Cycle 25. The Space-weather HMI Active Region Patches (SHARPs) have been identified automatically by \citet{bobra:2014} and for MDI (SMARPs) by \citet{bobra:2021}. The classification of an AR as a $\beta$-region is done by the National Oceanic and Atmospheric Administration (NOAA) on a daily basis and recorded in the Solar Region Summary, see \url{https://www.swpc.noaa.gov/products/solar-region-summary}, with that information imported and stored in the JSOC2 DRMS series su\_rsb.NOAA\_ActiveRegions. 
The HMI instrument measures the radial magnetic field ($B_{r}$) value, but MDI determines only the line-of-sight ($B_{los}$) component of the vector magnetic field. There are inherent differences between $B_{los}$ and $B_{r}$, and also differences between MDI and HMI $B_{los}$ \citep{liu:2012} due to instrumental design and performance differences. The tilt angles are calculated using the flux-weighted centroids of the positive and negative polarities such that Cycle 23 tilts are determined consistently using MDI data while Cycle 24 and 25 tilts are determined consistently using HMI data. We do not expect significant systematic differences in tilt angles measured by the two instruments, particularly since we restrict our analysis to regions not very close to the limbs.

\begin{figure}[!hb]
    \centering
    \includegraphics[width=0.65\textwidth]{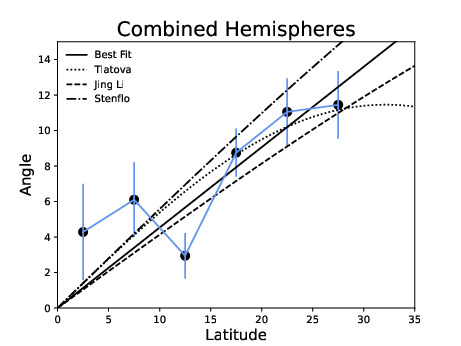}
    \caption{The best fit (solid black line) is fit to all $\gamma$ values for $\beta$-regions in Cycle 24. In order to combine the hemispheres onto a compact graph and for comparison with historical results, the zero point of the tilt angle was changed in each hemisphere such that 0$^\circ$ indicates the polarities are East-West aligned and both hemispheres have a positive angle increasing with latitude. The equation for the line of best fit, forced through the origin, is $\gamma$ = $(0.44 \pm 0.04)\theta$. Overplotted are the data points representing the average $\gamma$ values for Cycle 24 for the northern and southern hemispheres combined for $5^{\circ}$ latitude bins with a blue line connecting the data points. Error bars show the standard deviation of the mean for each bin.  Our fit is shown in context with the following results: \citet{tlatova:2018} $\gamma = $ 0.2~sin(2.8\,$\theta$) (in radians), \citet{Li:2018} $\gamma$ = 0.39\,$\theta$, and \citet{stenflo:2012} $\gamma$ = 32.1~sin\,$\theta$. The best fit line plotted here is the only time in this report that the line is forced through the origin to better compare our data to other studies that have y-intercepts of zero.}
    \label{fig:combined-hemispheres}
\end{figure}

AR patches that have multiple NOAA (National Oceanic and Atmospheric Administration) numbers are not included in the data set as they are not bipolar and are determined to be too complex to provide meaningful centroid calculations. It is common for ARs to form on the far side and rotate onto the front side of the Sun. These regions pose problems for studying the evolution of ARs as there is no way to determine what part of the AR lifetime is being observed. For this reason, ARs that did not emerge or exhibit a maximum amount of flux within 70$^\circ$ of central meridian are not included to ensure that all regions in the data set were observed to emerge on disc. 

The location of the leading and following polarities are determined by the flux-weighted centroids calculated in the MDI and HMI data and $\gamma$ values are defined as the tangent of the change in latitude and longitude between the flux-weighted centroids of the two polarities. $\gamma$ is calculated for a $0-360^\circ$ range defined with the negative polarity always at the origin and a 0$^\circ$  tilt indicating a completely vertical active region with the negative polarity to the north.  Angles increase counter-clockwise. A 90$^\circ$ tilt indicates a completely horizontal AR with negative leading polarity. Angles are recorded in this manner.  However, for most of the figures, the zero point of the tilt angle value is redefined such that a 0$^{\circ}$ tilt represents an AR with polarities aligned in an East-West direction, with the expected Joy's law tilt being positive in the northern hemisphere and negative in the northern hemisphere, and the tilt angle values translated accordingly in order to compare the Joy's law slope with those reported in past publications; the change in zero point is noted in the caption.

Anti-Hale regions are defined as active regions that don't obey Hale's law and therefore have the opposite leading polarity than expected. It is important to note that the tilt angles of anti-Hale regions are computed and recorded but are not used during the fitting of Joy's law or the determination of the $\bar{\gamma}$ values.

Figure~\ref{fig:tilt-vs-lat} shows the measured $\gamma$ as a function of latitude for $\beta$-type ARs in Cycle 23\,--\,25 measured at the time of peak flux when 100\% of the flux was emerged. In Cycle 24 (middle panel) the leading polarity in the northern hemisphere (latitude\,$>0^{\circ}$) is negative, therefore Hale regions are those with $\gamma$ between $0^{\circ}$\,--\,$180^{\circ}$. Northern hemisphere regions with $\gamma>180^{\circ}$ are anti-Hale.  
Northern ARs with $90^{\circ} < \gamma < 180^{\circ}$ are anti-Joy, but not anti-Hale. In Cycles 23 and 25 (top and bottom panels) the dominant leading polarity in the northern hemisphere switches to be positive, according to Hale's law. Therefore, Hale regions in the northern hemisphere have $180^{\circ} < \gamma < 360^{\circ}$.

AR tilt angles in our analysis are measured at three specific points in time, when 20\% of its flux has emerged, when 100\% of its flux has emerged, and at the time of central meridian crossing. Individual AR tilt angles are determined at a specific time and are not averaged in time. 

In order to analyze a relationship between the size of active regions and Joy's law, total unsigned flux is used as a reasonable indication of the size of the active region. The data are separated by hemisphere and size thresholds are determined in the following manner. 
We first record a single unsigned flux value for each AR at the time when 100\% of the AR's flux had emerged, i.e., the peak flux value. We then calculate the median flux value, $\Phi_{median}$, of this sample. After this, we divide the ARs into small, medium and large regions with small regions having a peak flux less than 2/3 $\times~\Phi_{median}$, medium regions having a peak flux greater than or equal to 2/3 but less than 4/3 $\times~\Phi_{median}$, and large regions having a peak flux greater than that. The resulting categories resulted in 384 small, 710 medium, and 502 large regions in Cycles 23 and 24. After the AR size classification, the ARs tilt values and Joy's law are examined when 100\% of its flux has emerged. We do not consider Cycle 25 because it is not yet complete and the small number statistics prevent a Joy's law slope from being determined once we separate by hemisphere and AR size. 

\begin{figure}[!hb]
    \centering
    \includegraphics[width=0.65\textwidth]{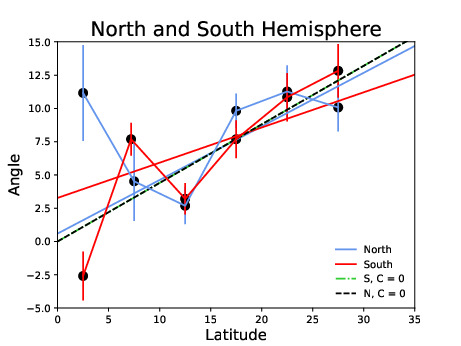}
    \caption{Average tilt angles from Cycle 24 in each hemisphere are shown separately and plotted in $5^{\circ}$ bins. Black dots show these averages connected by blue and red lines for the north and south hemispheres, respectively, with error bars calculated as the standard deviation of the mean. Best fit lines are fit to all data points in each data set, not the binned averages. Lines with intercepts are not forced through the origin (red and blue lines) in order to show the true best fit for the data, while the lines with intercepts forced through zero (dashed and dotted black lines) are shown for comparison. The black dashed line and green dashed and dotted line lie on top of each other and are virtually indistinguishable.}
    \label{fig:hemispheres}
\end{figure}

Tilt angles are plotted against latitude and a fit is performed using linear regressions with the form consistent with conventional Joy's Law fits:
\begin{equation} 
\gamma = m_{Joy}\theta + C
\end{equation}
where $\gamma$ is the tilt angle, $\theta$ is the latitude, $m_{Joy}$ is the slope, and $C$ is the y-intercept. $C=0$ if the fit is forced through the origin. The best fit is determined using all data points in all cases. Only one fit for Joy's law in Cycle 24 was forced through the origin (Figure~\ref{fig:combined-hemispheres}) in order to compare our results to previous studies in which the fit was forced through the origin motivated by the idea that the Coriolis force should tend towards zero at the equator. All other plots and fits were not forced through the origin. The average tilt angle determined for latitude bins of 5$^{\circ}$ width is overplotted on the Joy's law fits for visual purposes only.  Error bars on the binned data averages represent the standard error of the mean. The uncertainty of the slope and y-intercept of the fit are also reported. 

\section{Results}
\label{sec:results}

Figure \ref{fig:combined-hemispheres} shows the trend of $\gamma$ as a function of latitude for all $\beta$-regions at their maximum unsigned magnetic flux in Cycle 24. The northern (southern) hemisphere tilt values in the $0-360^{\circ}$ range were clustered near $90^{\circ}$ ($270^{\circ}$) for Cycle 24.  The black data points connected by the blue line in Figure~\ref{fig:combined-hemispheres} represent $\gamma$ values averaged in $5^{\circ}$ latitude bins. The best fit (solid black line) was fit to all of the scatter points in the data set, not just the binned averages shown on the graph. The best linear fit to Cycle 24 data, with both hemispheres combined and the fit forced through the origin, was found to be $\gamma$= (0.44 $\pm$ 0.04)$\theta$, see Figure~\ref{fig:combined-hemispheres}.

\subsection{Fits Separated by Hemisphere}
\label{sec:hem}

The $\gamma$ of ARs sampled at the time of their maximum (100\%) flux are divided into northern and southern hemispheres and analyzed individually in order to observe the differences in activity as shown in Figure \ref{fig:hemispheres}. In order to show both hemispheres in a compact graph, the zero point of the tilt angle was changed such that both hemispheres would have a positive angle increasing with latitude in accordance with Joy's law. To achieve this, we subtracted $\gamma$ from 90$^{\circ}$ in the northern hemisphere and subtracted 270$^{\circ}$ from $\gamma$ in the southern hemisphere. Linear regressions are used to fit the data with the form $\gamma = m_{Joy}\theta + C$.  Linear fits and the overplotted average $\gamma$ values from 5$^{\circ}$ latitude bins are shown for the northern and southern hemispheres following the procedure detailed earlier with anti-Hale regions being excluded. Results show slightly different linear fits, with the lowest latitude bins having distinctly different tilts. Hemispheric fits that were forced through the origin are overplotted with dashed and dotted lines. 

The northern hemisphere in Figure \ref{fig:hemispheres} shows a clear downturn in the 25 -- 30$^{\circ}$ band, also observed by \citet{tlatova:2018} and seen in Figure~\ref{fig:combined-hemispheres}. The best fit lines for the two data sets also appear to be different, although these differences cannot be proven to be statistically significant due to the lower sample size after separating the hemispheres, see Table 1 entries for 100\% flux for slope ($m_{Joy}$), intercept (C), uncertainties ($\sigma_m$ and $\sigma_C$), average tilt angle $\bar{\gamma}$ and associated uncertainty ($\sigma_{\bar{\gamma}}$), median tilt angle ($\gamma_{median}$) and the number of data points (N).  These Cycle 24 Joy's law parameter entries for 100\% flux emerged shown in Table 1 are repeated in Table \ref{tab:table-2} for Cycle 24.

It is worth noting that the mean tilt values in the 10 -- 15$^{\circ}$ latitude bins in Figures \ref{fig:combined-hemispheres} and \ref{fig:hemispheres} decrease in magnitude and therefore do not fit the trend of Joy's law. 
\begin{figure}[!ht]
    \centering
    \includegraphics[width=0.65\textwidth]{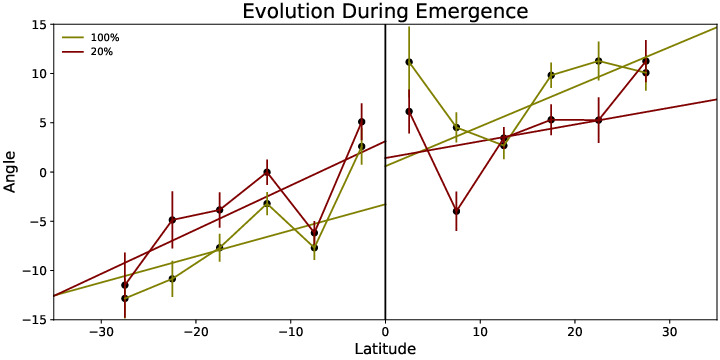}
    \caption{Tilt angles are calculated for all $\beta$ regions in Cycle 24 at times when 100\% and 20\% of their maximum unsigned flux has emerged. The two hemispheres are analyzed separately. Black data points represent average $\gamma$ in  $5^{\circ}$ latitude bins. Green and brown lines connect the 100\% and 20\% values, respectively, with error bars representing standard error of the mean for each averaged bin.The green lines for the tilts in the northern (southern) hemisphere in this plot correlate with the blue (red) lines in Figure \ref{fig:hemispheres} with the exception that the Southern hemispheric values in Figure \ref{fig:hemispheres} differ by a factor of $-1$ so that both hemispheric tilt angle values can be easily compared in Figure \ref{fig:hemispheres} and a larger range of values, $-90^{\circ}$ -- $90^{\circ}$, can be represented in this plot.}
    \label{fig:evol}
\end{figure}

\subsection{Joy's Law Dependent On Phase of Emergence}
\label{sec:phase}

To investigate whether $\bar{\gamma}$ changes during AR emergence, we sample AR evolution at times when different amounts of the maximum unsigned magnetic flux has emerged. Figure \ref{fig:evol} plots Joy's law for $\beta$-regions in Cycle 24 when 20\% and 100\% of the maximum unsigned flux for each region has emerged.  The hemispheres are separated. Table \ref{tab:table-2} shows all values and uncertainties associated with fitting Joy's law as shown in Figure \ref{fig:evol}.

As seen in Figure \ref{fig:evol} in the northern hemisphere, the Joy's Law slope increases from 0.27 to 0.40 as the unsigned magnetic flux completes its emergence. The southern hemisphere shows a reversed trend so the increase or decrease in slope is not consistent, and the change in the slope is not statistically significant due to the uncertainties, see $\sigma_m$ values in Table \ref{tab:table-2}. 

The significant result is that the average tilt angle, $\bar{\gamma}$ increases more than 3$^{\circ}$ between the time 20\% of the flux has emerged and when 100\% of the flux has emerged (see Table~1). 
This is true in both hemispheres and in the combined hemispheric data.

\begin{table}[!ht]
\begin{center}
\begin{tabular}{|c|c|c|c|c|c|c|c|c|c|} \hline
Flux Emerged (\%) & Hemisphere & $m_{Joy}$ & $\sigma_m$ & C & $\sigma_C$ & $\bar{\gamma}$ & $\sigma_{\bar{\gamma}}$& $\gamma_{median}$ & N \\ \cline{1-10}
\multirow{3}{*}{100\%} & North & 0.40 & 0.14 & 0.59 & 2.06 & 6.04 & 0.90 & 6.85 & 480\\ \cline{2-10}
& South & 0.26 & 0.14 & -3.27 & 2.42 & 7.42 & 0.98 & 6.61 & 399 \\ \cline{2-10}
& Combined & 0.34 & 0.10 & 1.68 & 1.56 & 6.67 & 0.66 & 6.79 & 879\\ \cline{1-10}
\multirow{3}{*}{20\%} & North & 0.27 & 0.16 & -0.89 & 2.41 & 2.74 & 1.05 & 2.50 & 426\\ \cline{2-10}
& South & 0.45 & 0.15 & 3.12 & 2.60 & 4.51 & 1.06 & 5.80 & 369\\ \cline{2-10}
& Combined & 0.36 & 0.11 & -1.97 & 1.75 & 3.27 & 0.75 & 3.30 & 795\\ \cline{1-10}
\end{tabular} 
\caption{Parameters from the fits to Joy's law for tilt angles in Cycle 24 for active regions that have emerged 100\% and 20\% of maximum flux. The fits are performed separately for each hemisphere and then a combined fit is done. From left to right the columns are the percentage of emerged flux, hemisphere (or combined data from both hemispheres), slope of the Joy's law fit $m_{Joy}$, uncertainty on the slope $\sigma_m$, y-intercept C, uncertainty of the y-intercept $\sigma_C$, the average tilt angle $\bar{\gamma}$, uncertainty of the tilt angle $\sigma_{\bar{\gamma}}$, the median tilt angle $\gamma_{median}$, and the number of points in each data set N. Uncertainties are calculated as the standard error of the mean.}
\end{center}
\label{tab:table-1} 
\end{table}

\subsection{Joy's Law Dependency on Size of AR}
\label{sec:size}

\begin{figure}[!ht]
    \centering
    \includegraphics[width=0.65\textwidth]{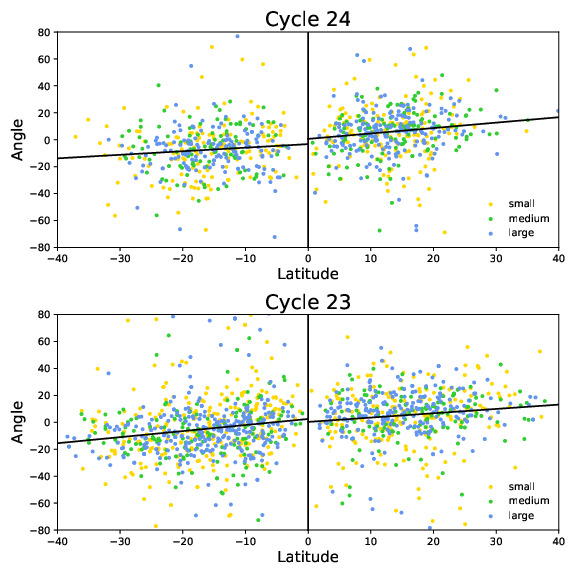}
    \caption{Tilt angles of each active region in Cycles 23 and 24 are plotted at the time of the maximum unsigned flux. Yellow, green, and blue points show results for small, medium and large active regions. The best fit line shown in black is a fit to all of the points. We find no significant difference between the best fit lines of different sized active regions and therefore do not show these lines.The fits for Cycle 24 are the same as shown in Fig. \ref{fig:evol}.
    }
    \label{fig:AR-size}
\end{figure}

AR $\gamma$ values for both Cycle 23 and Cycle 24 are categorized by size and shown as a function of latitude in Figure~\ref{fig:AR-size} in different colors for each size and separated by hemisphere in order to determine if $\gamma$ values and Joy's law fits are a function of the AR size. Size classifications into small, medium and large regions are made following the procedure detailed in Section 2 using the median of the peak flux values ($\Phi_{median}$) for each hemisphere and cycle. $\Phi_{median}$ values of the south and north hemispheres in Cycle 24 were $8.94\times10^{21}$ and $9.26\times10^{21}$~Mx, respectively. $\Phi_{median}$ of the south and north hemispheres in Cycle 23 was $8.51\times10^{21}$ and $9.02\times10^{21}$ Mx, respectively.  The average time for the small, medium and large regions to emerge from 20\% flux to 100\% flux was found to be 2.01, 4.86, and 7.42 days, respectively. 

There is no strong correlation between the size of the active regions and the magnitude of their average tilt angles. While the intercepts and slopes vary greatly, they all have large uncertainties due to the high scatter of the data set. We therefore conclude that there is no statistically significant trend and only plot the best fit line (separated by hemispheres) to all data points for Joy's law for all sizes of ARs, see the black lines in Figure~\ref{fig:AR-size}.  Figure~\ref{fig:AR-size} shows the large scatter of the tilt angles of active regions of different sizes in the yellow, green and blue data points representing the small, medium and large ARs. 

\subsection{Comparison Between Solar Cycles}
\label{sec:cycles}

\begin{figure}[!ht]
    \centering
    \includegraphics[width=0.65\textwidth]{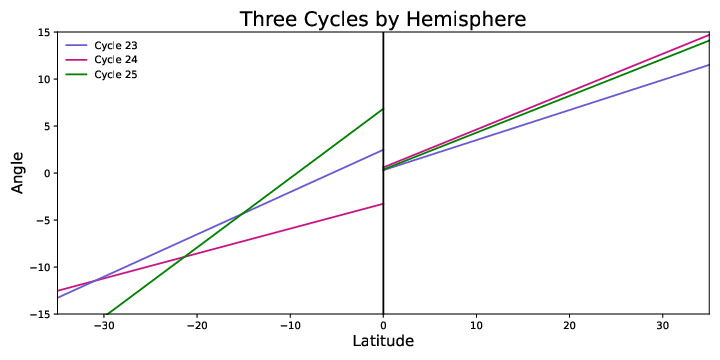}
    \caption{Best fit lines for all $\beta$-regions in the northern and southern hemispheres in Cycles 23 (lilac), 24 (magenta) and the early portion of 25 (green). Lines are fit to all data points in each data set and are not forced through the origin.  }
    \label{fig:hemispheres-by-cycle}
\end{figure}

Figure \ref{fig:hemispheres-by-cycle} shows best fit lines for Cycles 23, 24 and the beginning of 25 in the two hemispheres when the AR flux is 100\% emerged. The northern hemisphere is very consistent and the differences in the slopes and intercepts are statistically insignificant from cycle to cycle. The southern hemisphere is much more variable, with significant differences in the slopes between Cycles 23 and 24, see the difference between the blue and red lines in the southern hemisphere in Figure \ref{fig:evol}. The Cycle 25 data is incomplete with only 20-30\% of the number of data points in the other two cycles. However, the difference in slope between the northern and southern hemispheres in Cycle 25 is quite large even this early in the cycle, see the green line in Figure \ref{fig:evol}. The fitted parameters and their associated uncertainties are presented in Table \ref{tab:table-2}.

The $\bar{\gamma}$ values for the hemisphere-combined data are significantly different between Cycle 23 (5.11$^{\circ}$ $\pm$ 0.61) and Cycle 24 (6.67$^{\circ}$ $\pm$ 0.66). The Southern hemisphere contributes most of this difference with $\bar{\gamma}$ values of Cycle 24 being 2.55$^{\circ}$ higher than those of Cycle 23, see Table \ref{tab:table-2}.  

\begin{figure}[!ht]
    \centering
    \includegraphics[width=0.65\textwidth]{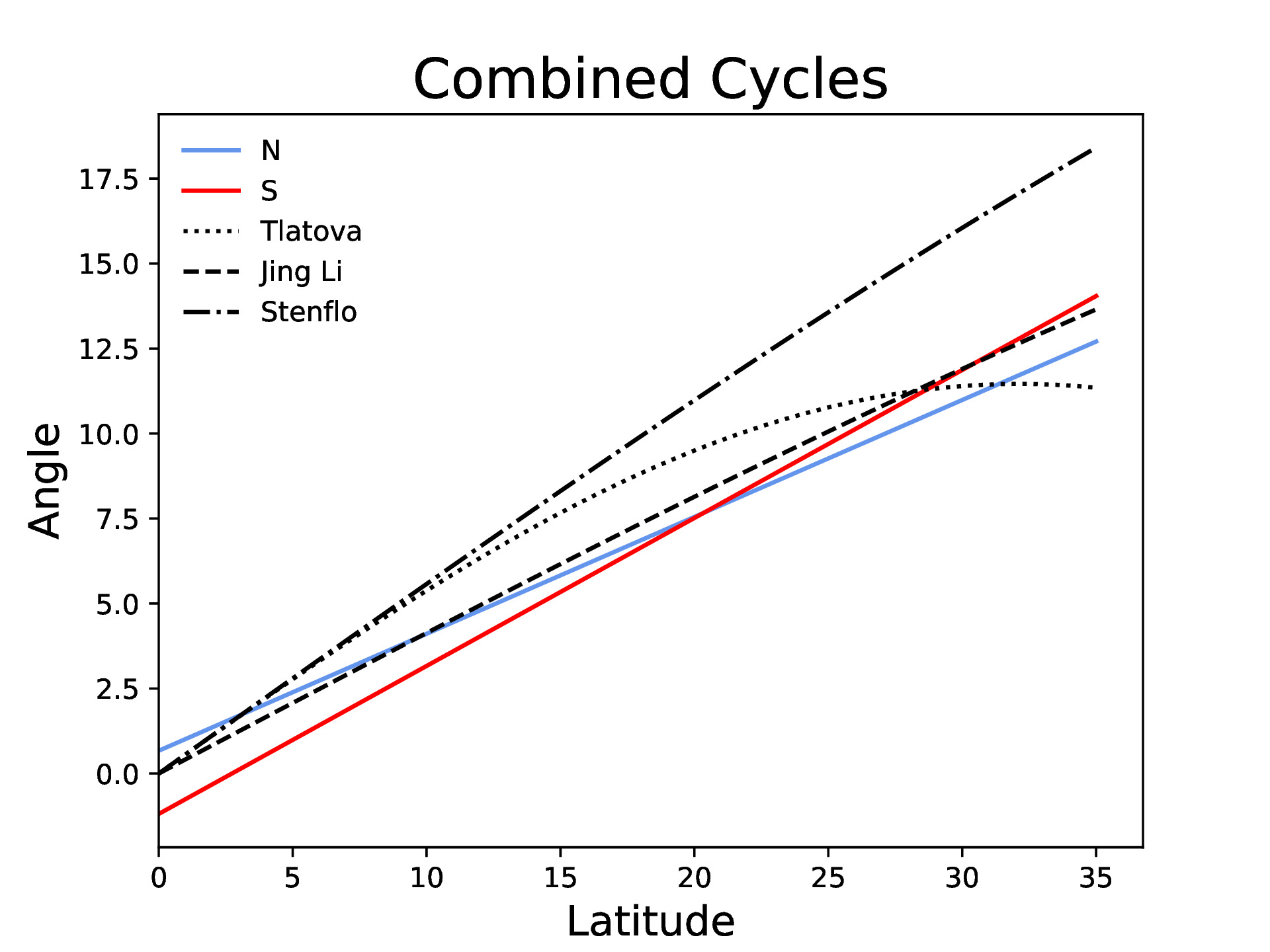}
    \caption{The best fit line for all $\beta$-regions in each hemisphere for the combined Cycles 23, 24, and the beginning of Cycle 25. The best fit lines are not forced through the origin, however it is worth noting that the intercepts are very small compared to those from fits for each separate cycle.}
    \label{fig:combined-cycles}
\end{figure}

To compare the averages of the hemispheric behavior for the 27 years, all data from the three cycles are combined into two data sets, one for each hemisphere, and a linear regression is fit to both as shown in Figure \ref{fig:combined-cycles}. The southern hemisphere had an overall higher slope and a negative intercept with the northern hemisphere having a smaller slope and positive intercept. Both the northern and southern intercepts, however, are statistically consistent with 0. Comparing these results to other studies reveals similar results within the bounds of uncertainty. The fits to the combined hemispheric data for 27 years can be found in the last two rows of Table \ref{tab:table-2}.  We find these results to be in good agreement with \citet{tlatova:2018}, \citet{Li:2018}, and \citet{stenflo:2012}.


\begin{table}[!hb]
\begin{center}
\begin{tabular}{|c|c|c|c|c|c|c|c|c|c|} \hline
    Cycle & Hemisphere & $m_{Joy}$ & $\sigma_m$ & C & $\sigma_C$ & $\bar{\gamma}$ & $\sigma_{\bar{\gamma}}$& $\gamma_{median}$ & N \\ \cline{1-10}
    \multirow{2}{*}{Cycle 25} & North & 0.39 & 0.26 & 0.37 & 5.16 & 7.80 & 1.56 & 10.33 & 150\\ \cline{2-10}
    & South & 0.74 & 0.17 & -6.83 & 3.63 & 8.19 & 1.31 & 8.41 & 154 \\ \cline{2-10}
    & Combined & 0.60 & 0.15 & -3.80 & 3.04 & 8.00 & 1.02 & 9.49 & 304\\ \cline{1-10}
    \multirow{2}{*}{Cycle 24} & North & 0.40 & 0.14 & 0.59 & 2.06 & 6.04 & 0.90 & 6.85 & 480\\ \cline{2-10}
    & South & 0.26 & 0.14 & -3.27 & 2.42 & 7.42 & 0.98 & 6.61 & 399 \\ \cline{2-10}
    & Combined & 0.34 & 0.10 & 1.68 & 1.56 & 6.67 & 0.66 & 6.79 & 879\\ \cline{1-10}
    \multirow{2}{*}{Cycle 23} & North & 0.32 & 0.11 & 0.30 & 1.93 & 5.41 & 0.88 &  7.00 & 593 \\ \cline{2-10}
    & South & 0.45 & 0.10 & -2.14 & 1.80 & 4.87 & 0.83 & 6.36 & 763 \\ \cline{2-10}
    & Combined & 0.39 & 0.07 & -1.26 & 1.32 & 5.11 & 0.61 & 6.68 & 1356\\ \cline{1-10}
    \multirow{2}{*}{Combined Cycles} & North & 0.34 & 0.08 & 0.70 & 1.33 & 5.95 &  0.59 & 7.28 & 1223 \\ \cline{2-10}
    & South & 0.43 & 0.07 & -1.19 & 1.34 & 6.05 & 0.59 & 6.96 & 1316\\ \cline{1-10}
\end{tabular}
\caption{Fits to Joy's law for tilt angles for Cycles 23, 24, and 25. The fits are performed separately for each hemisphere and a combined fit is shown. From left to right the columns are the percentage of emerged flux, hemisphere (or combined data from both hemispheres), slope of the Joy's law fit $m_{Joy}$, uncertainty on the slope $\sigma_m$, y-intercept C, uncertainty of the y-intercept $\sigma_C$, the average tilt angle $\bar{\gamma}$, uncertainty of the tilt angle $\sigma_{\bar{\gamma}}$, the median tilt angle $\gamma_{median}$, and the number of points in each data set N. Uncertainties are calculated as the standard error of the mean.}
\label{tab:table-2} 
\end{center}
\end{table}

\subsection{Significance of the Time of Sampling}
\label{sec:cm}

Tilt angles measured at the time of maximum flux of each active region and from the time of central meridian crossing are compared to determine if a significant difference exists in the average tilt angles or Joy's law fit that depends on the sampling. The data from central meridian crossing was taken from the SPEAR (Solar Photospheric and Ephemeral Active Region) catalogue, see \url{http://hmi.stanford.edu/hminuggets/?p=3730}. 

Figure \ref{fig:cm-maxflux} shows the average 5$^{\circ}$ bins for the data points from both data sets separated by hemisphere and the linear fit done to all data points (not the binned data). There is no significant difference in the linear regression slopes and intercepts in the data from the northern hemisphere and in the intercepts from the southern hemisphere, see Table \ref{tab:table-3}. The $\bar{\gamma}$ values resulting from sampling at the time of maximum flux and central meridian are also statistically the same. The average difference between maximum flux and central meridian was $0.82^{\circ}$ in the northern hemisphere and $3.83^{\circ}$ in the southern hemisphere. 

\begin{table}[!hb]
\begin{center}
\begin{tabular}{|c|c|c|c|c|c|c|c|c|c|} \hline
    Time Measured & Hemisphere & $m_{Joy}$ & $\sigma_m$ & C & $\sigma_C$ & $\bar{\gamma}$ & $\sigma_{\bar{\gamma}}$ & $\gamma_{median}$ & N \\ \cline{1-10}
    \multirow{3}{*}{Maximum Flux} & North & 0.66 & 0.31 & 0.25 & 4.32 & 9.33 & 1.74 & 7.78 & 111 \\ \cline{2-10}
    & South & 0.22 & 0.32 & -2.44 & 5.42 & 7.22 & 1.77 & 7.12 & 93 \\ \cline{2-10}
    & Combined & 0.38 & 0.21 & 2.19 & 3.38 & 7.80 & 1.35 & 6.62 & 204 \\ \cline{1-10}
    \multirow{3}{*}{Central Meridian} & North & 0.45 & 0.29 & 2.16 & 4.20 & 8.25 & 1.66 & 7.56 & 112\\ \cline{2-10}
    & South & 0.40 & 0.26 & -0.65 & 4.53 & 7.22 & 1.77 & 7.12 & 90\\ \cline{2-10}
    & Combined & 0.39 & 0.19 & 2.04 & 3.01 & 7.79 & 1.21 & 7.36 & 202 \\ \cline{1-10}
\end{tabular}
\caption{The parameters of the best fit line for Joy's law for tilt angles of ARs in Cycle 24 as sampled at their time of maximum unsigned flux and at their central meridian crossing times. Restrictions on the data set require that the region be a $\beta$-type AR at both the time of maximum flux and at central meridian crossing, leading to a smaller data set than that used for previous tables. Regions were also required to emerge within $70^{\circ}$ of central meridian. From left to right the columns are the percentage of emerged flux, hemisphere (or combined data from both hemispheres), slope of the Joy's law fit $m_{Joy}$, uncertainty on the slope $\sigma_m$, y-intercept C, uncertainty of the y-intercept $\sigma_C$, the average tilt angle $\bar{\gamma}$, uncertainty of the tilt angle $\sigma_{\bar{\gamma}}$, the median tilt angle $\gamma_{median}$, and the number of points in each data set N.}
\label{tab:table-3}
\end{center}
\end{table}

\section{Discussion}
\label{sec:discussion}

Joy’s law is a statistical law and the large scatter means that the results are sensitive to sample size and choices in fitting techniques. If ARs are only sampled once and the hemispheres are separated, it is difficult to find a statistical difference between the slopes and intercept of Joy's law, especially in Cycle 24 as the number of data points are on the order of $\sim$450. 

The most significant result from this work is that the $\bar{\gamma}$ values in the combined hemispheric data increase from 3.27$\pm$0.75$^{\circ}$ when 20\% of the flux has emerged in an AR to 6.67$\pm$0.66$^{\circ}$ when 100\% of the flux has emerged (Table 1).  This confirms the results of \citet{schunker:2020} who found that ARs initially emerge nearly East-West aligned and the tilt angles become larger, on average, as more flux emerges.  The individual hemispheres exhibit the same trend with an approximately 3$^{\circ}$ increase in $\bar{\gamma}$ values during emergence. \citet{stenflo:2012} also report an evolution towards the standard Joy's law tilt over the course of an AR lifetime.  

\begin{figure}[!ht]
    \centering
    \includegraphics[width=0.65\textwidth]{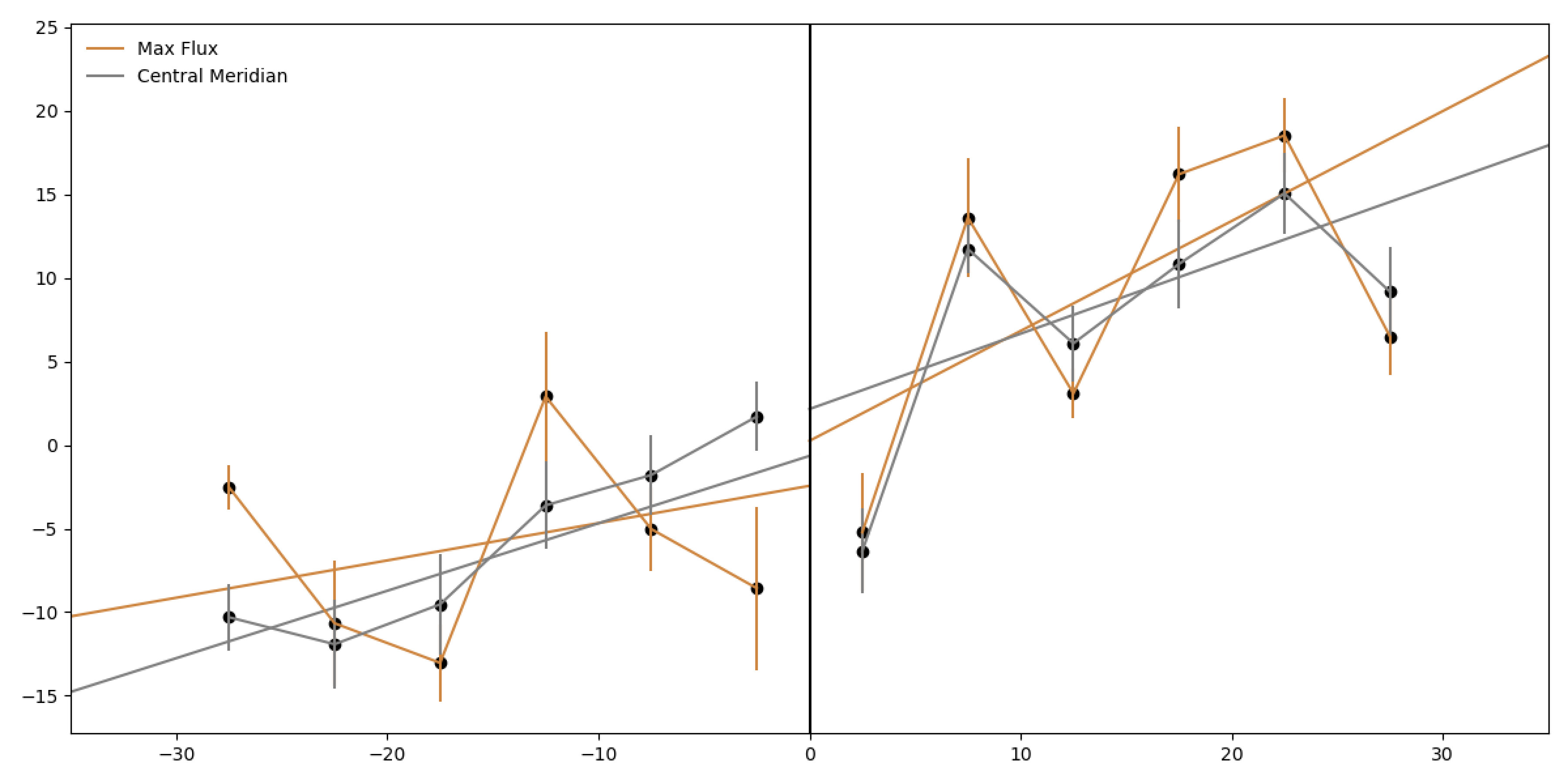}
    \caption{Comparison of Joy's Law in Cycle 24 determined by sampling the ARs at the time of maximum flux versus the central meridian crossing. Best fit lines are fit to all $\beta$ regions and are not forced through the origin and fit to all $\beta$-type regions. Overlaid points represent average $\gamma$ values from 5$^{\circ}$ bins in latitude. }
    \label{fig:cm-maxflux}
\end{figure}

One hypothesis we pursued, that ended up being a null result, was that fitting Joy's law at the time of maximum flux of an AR would be statistically different than the fit at the time of central meridian crossing. This is not what we discovered (see Table \ref{tab:table-3} and Figure \ref{fig:cm-maxflux}). Joy's law fits are similar for data sampled at the time of AR maximum flux and for those sampled at the time of central meridian crossing in this data. This is probably because the scatter in the tilt angles was quite high, which was due to limiting the sample size by separating hemispheres and to requiring that the ARs emerge on the front side of the disk, resulting in a sample of only $\sim$100-200 regions. The central meridian sampled data does not take into account the stage of development of the active region, therefore it can be assumed that it samples regions at all stages of emergence and decay. 

Our best linear fit for Cycle 24 data with both hemispheres combined and forcing the fit through the origin is $\gamma$=(0.44 $\pm$ 0.04)\,$\theta$, see Figure~\ref{fig:combined-hemispheres}. This is consistent with \citet{Li:2018} reporting a best fit of $\gamma = (0.39 \pm 0.06)\,\theta - (0.66 \pm 1.00)$ and other fits to data that contain magnetic polarity information.  Notably, the tilt angles in HMI data do not increase linearly with latitude but plateau above 20$^{\circ}$. This agrees with the finding of \citep{tlatova:2018}, also shown in Figure ~\ref{fig:combined-hemispheres}, such that the AR tilt angles at latitudes higher than the 25-30$^{\circ}$ bin do not increase. Figures 1 and 3 in \citet{wang:2015} showing data from Cycles 21 - 23, both median and mean tilts and white-light and magnetogram data, also show that tilt values decline above 25 - 30$^{\circ}$ latitude.  Since $\gamma$ does not increase linearly for latitudes higher than 25$^{\circ}$, the form of Joy's law proposed by \citet{tlatova:2018} $\gamma = 0.2\,sin(2.8\,\theta)$ is sensible.  If adopted by modelers, it would ensure that any modeled ARs that emerge at high latitudes do not have an unrealistically large tilt that unduly influences the next cycle.  

The fits shown in Figure~\ref{fig:hemispheres} emphasize the difference between forcing the linear regressions through the origin and allowing the fits to have a y-intercept. Many other studies, including \citet{stenflo:2012}, \citet{Li:2018} and \citet{tlatova:2018}, computed Joy's Law fits with linear regressions forced through the origin under the assumption that tilt angles at the equator should be $0^{\circ}$. However, in doing so, some information is lost in the process. When forcing the best fits through the origin (black and grey dotted line), they both have the same slope. This leads to the conclusion that the two hemispheres are quite symmetrical, as assumed by \citet{stenflo:2012} and many others. However, when plotting the linear regressions without forcing the lines through the origin (blue and red), the slopes are different slopes and have different intercepts. While these slopes are not statistically significant due to the large scatter in the data, we suggest that these results reflect a difference in hemispheric activity.

An interesting feature reported by \cite{tlatova:2018} is that at 0$^{\circ}$ latitude, odd cycles have positive offsets, i.e. a positive tilt angle value, and even cycles have negative offsets, i.e. a negative tilt angle value, up through Cycle 22. We did not observe a similar effect, although the difference may be due to differences in fitting:  \citet{tlatova:2018} fit the latitudinally binned data that included multiple entries for a given AR, while our procedure only samples each AR once and fits all data points, not the binned averages. However, this may be an interesting point for further investigation, especially if one can concentrate only on the lower-latitude ARs that appear to be the regions contributing to the offset.  

Previous research has reported that $\gamma$ and Joy's law slopes are not dependent on the amount of flux contained within the AR or the size of the AR \citep{fisher:1995, stenflo:2012}. Our attempt to distinguish a dependency of the tilt or Joy's law on the size of ARs in MDI and HMI also led to a null result. In contrast, \citet{sreedevi:2024, sreedevi:2023} reports an increasing trend in tilt angles as a function of flux of the active regions when excluding the highest flux regions and averaging over two solar cycles and both hemispheres.

The fact that $\bar{\gamma}$ values are higher in Cycle 24 (6.67$^{\circ}$ $\pm$ 0.66) than Cycle 23 (5.11$^{\circ}$ $\pm$ 0.61) fits into the theory that tilt angles are a feedback mechanism to limit the runaway growth of a solar cycle \citep{jiang:2020} and that there is a tendency for weaker cycles to produce larger tilt angles than stronger cycles \citep{dasi-espuig:2010}.

A relatively recently studied feature of ARs that may be influential in the determination of $\gamma$ is the ``magnetic tongue."  At the very early stage of emergence, the twist of the magnetic field on the flux tube may influence the measured tilt angle as the azimuthal component of the field is projected onto the vertical, see the work of \citet{poisson:2016} on magnetic tongues (or tails). Preliminary studies show that identifying and removing these magnetic tongues from the AR data prior to calculating the tilt angle may lower the scatter in the values \citep{poisson:2020}. 

HMI and MDI have provided an abundance of magnetic field data of solar ARs for the past 27 years.  Tilt angles and Joy’s law have been studied using these observations, as well as data that have come before. As yet, the community has not reached a consensus as to the underlying mechanism that imparts the tilts. However, increasingly detailed and creative studies are being conducted and with them, and there is hope for more clarity as to the role that the Coriolis effect plays versus other mechanisms that may impart AR tilt angles.

\bibliography{joy}{}
\bibliographystyle{aasjournal}
\end{document}